\documentstyle[aps,prd,epsfig]{revtex}
\begin{document}

\tightenlines

\title{Effective Lagrangian description of the lepton flavor violating
decays $Z \to \ell_i^{\mp} \ell_j^{\pm}$}
\author{A. Flores-Tlalpa$^a$, J. M. Hern\' andez$^b$,
G. Tavares-Velasco$^a$ and J. J. Toscano$^b$}
\address{$^a$Departamento de F\'\i sica, CINVESTAV, Apartado Postal 14-740,
07000, M\' exico, D. F., M\' exico}
\address{$^b$Facultad de Ciencias F\'\i sico Matem\' aticas, Benem\' erita
Universidad Aut\' onoma de Puebla, Apartado Postal 1152, 72000,
Puebla, Pue., M\' exico}

\date{\today}
\maketitle

\begin{abstract}
A comprehensive analysis of the lepton flavor violating (LFV)
decays $Z \to \ell_i^{\mp} \ell_j^{\pm}$ is presented within the
effective Lagrangian approach. Both the decoupling and
nondecoupling scenarios are explored. The experimental constraints
from $\ell_i \to \ell_j \ell_k \bar{\ell}_k$ and $\ell_i \to
\ell_j \gamma$  as well as some relationships arising from the
gauge invariance of the effective Lagrangian are used to put
constraints on $Z \to \ell_i^{\mp} \ell_j^{\pm}$. It is found that
while current experimental data impose very strong constraints on
$Z \to \mu^{\mp} e^{\pm}$, the channel $Z \to \tau^{\mp}
\ell^{\pm}$ still may be at the reach of the planned TESLA
collider.

\end{abstract}

\draft 

\section{Introduction}
Recent neutrino experimental data, such as those coming from
Super-Kamiokande \cite{kk}, have shown evidences of atmospheric
and solar neutrino oscillations. This class of effects point to
physics beyond the standard model (SM) and have immediate
consequences on some sectors of the theory. For instance, the
conservation of lepton number and lepton flavor can not be taken
for granted anymore, as in the SM with its massless neutrinos.
Clearly, some lepton flavor violating (LFV) processes such as $Z
\to \ell_i^{\mp} \ell_j^{\pm}$ ($\ell_{i}=e,\,\mu,\,\tau$) may
occur and be observable at the future particle colliders. The
neutrino experimental data have thus renewed the interest in LFV
transitions. Moreover, the prospect of the $e^+e^-$ TeV energy
superconducting linear accelerator (TESLA) with its Giga-$Z$
option \cite{tesla1} opens up the possibility of studying at a
near future some LFV $Z$ boson decays, which might be at the reach
of that collider \cite{illana}. Currently, the best direct
experimental bounds on the $Z \to \ell_i^{\mp} \ell_j^{\pm}$
rates, obtained by the search at LEP-I, are \cite{pdg}

\begin{mathletters}
\begin{equation}
{\mathrm{BR}}\left(Z \to e^{\mp} \mu^{\pm}\right)< 1.7 \times
10^{-6},
\end{equation}
\begin{equation}
{\mathrm{BR}}\left(Z \to e^{\mp} \tau^{\pm}\right)< 9.8 \times
10^{-6},
\end{equation}
\begin{equation}
{\mathrm{BR}}\left(Z \to\mu^{\mp} \tau^{\pm}\right)<1.2 \times
10^{-5},
\end{equation}
\end{mathletters}

\noindent whereas the expectations at TESLA are \cite{tesla1}
\begin{mathletters}
\begin{equation}
{\mathrm{BR}}\left(Z \to e^{\mp} \mu^{\pm}\right)< 2.0 \times
10^{-9},
\end{equation}
\begin{equation}
{\mathrm{BR}}\left(Z \to e^{\mp} \tau^{\pm}\right)< (1.3-6.25)
\times 10^{-8},
\end{equation}
\begin{equation}
{\mathrm{BR}}\left(Z \to\mu^{\mp} \tau^{\pm}\right)<(0.44-2.2)
\times 10^{-8}.
\end{equation}
\end{mathletters}

In order to disentangle the origin of any possible LFV effect,
TESLA expectations must be confronted with the predictions of the
diverse available models. Considerable work has been done along
these lines, but here we will only refer to the most recent studies.
For instance, the authors of Ref. \cite{illana} reviewed diverse
scenarios that enlarge the SM by just adding massive neutrinos. It
turns out that, after considering the most recent experimental
data for neutrino masses, one can have at most
${\mathrm{BR}}\left(Z\to \ell_i^{\mp}\ell_j^{\pm}\right) \sim
10^{-54}$ for light neutrinos. On the other hand,  $Z\to
\ell_i^{\mp}\ell_j^{\pm}$ might be at the reach of TESLA in some
models with heavy neutrinos whose mass is of the order of
$200-1000$ GeV. This decay has also been studied within the
general two Higgs Doublet Model \cite{iltan}. It was found that
the channel $Z \to \mu^{\mp} e^{\pm}$ is the only one that may be
at the reach of TESLA. Studies within the Zee Model \cite{ghosal}
and theories with a heavy $Z^\prime$ boson with family
non-universal couplings \cite{langacker} gave results that are far
from the experimental limits. Further works have been realized
within other models, such as supersymmetry, leptoquark theories,
left-right symmetric models, etc. \cite{ZLFV}.

All of the aforementioned studies have focused on specific models,
which share the common feature of being of a weakly coupled
nature, i.e. when the masses of the heavy particles become large
they decouple from low energy physics. Therefore, it is convenient
to take a more general approach that allows us to make a
model-independent analysis. We will consider thus the effective
Lagrangian approach (ELA), which is suitable for this purpose. In
particular, the ELA has been extensively used to study some
quantities that are forbidden or highly suppressed within the SM.
In this approach there are two well-motivated schemes to
parametrize virtual effects of particles lying beyond the Fermi
scale via effective operators involving only the SM fields, namely
the linear and nonlinear realizations of the electroweak group.

In the linear realization or decoupling scenario \cite{buch} it is
assumed that the spontaneous symmetry breaking (SSB) of the
electroweak group takes place in the usual way, thereby implying
the existence of at least one physical Higgs boson. In addition,
the light particles (the SM ones) fill out multiplets of
$SU_L(2)\times U_Y(1)$. Although one can only expect marginal
contributions from the heavy fields to low energy physics, there
are indeed some processes in which the new physics effects may
compete with the SM ones, such as those involving  flavor-changing
neutral current (FCNC) and LFV transitions \cite{DT}. The latter
are forbidden in the SM at any order of perturbation theory. The
decoupling scenario is suitable to parametrize any virtual effect
arising from a fundamental gauge theory that is assumed to be
renormalizable and of a weakly coupled nature. This hypothesis is
fundamental to establish a hierarchy among those operators of a
particular dimension: gauge invariance allows us to infer the
order at which the effective operators may be generated in
perturbation theory \cite{AEW}. In particular, a loop-generated operator
is suppressed by a factor of $(4\pi)^{-2}$ with respect to a
tree-level induced one. Throughout this work we will make
systematic use of this fact when studying LFV processes mediated
by the $Z$ boson.

As to the nonlinear realization or nondecoupling scenario
\cite{NL}, in this case it is assumed that the low-energy processes are
affected by unknown residual strong-dynamics effects. In this
effective (chiral) theory, the SSB of the electroweak group is
accomplished by introducing a unitary matrix field $U$ that
replaces the SM doublet. It is also assumed that the physical
Higgs boson either is very heavy or does not exist at all. The
scalar sector is comprised only by Goldstone bosons that define
the $U$ field, which in turn transforms nonlinearly under the
$SU_L(2)\times U_Y(1)$ group. In the unitary gauge, where the
Goldstone bosons are absent, we have that $U=1$ and the chiral
Lagrangian reproduces the SM without the Higgs field. Due to the
fact that a strongly interacting regimen implies that loop effects
can be as important as the tree-level ones, one can not establish
a priori what operators are the most relevant. We will bear this
fact in mind when we discuss the general structure of the $Z\ell_i
\ell_j$ couplings within the nondecoupling scenario.

Our main aim is thus to present a model independent study of the
LFV decay $Z \to \ell_i^{\mp} \ell_j^{\pm}$ within the ELA. We
will make general predictions for the respective rates in both the
decoupling and nondecoupling scenarios. Further, the impact on
this decay of the experimental constraints on $\ell_i \to \ell_j
\ell_k \bar{\ell}_k$ and $\ell_i \to \ell_j \gamma$ will be
analyzed, and the expectations at the future TESLA collider
\cite{tesla}, running at the $Z$ peak (Giga-Z), will be discussed.

The rest of our presentation is organized as follows. In Sec.
\ref{dc} we consider the decoupling scenario and discuss the most
general structure of the $Z\ell_i \ell_j$ vertex. It is argued
that the contribution from the monopole structure $\gamma_\mu$
dominates over that from the dipole structure $\sigma_{\mu \nu}\,
k^\nu$: the latter can only arise at the one-loop level in any
renormalizable theory. Therefore, the most stringent bounds on
$Z\to \ell_i^{\mp}\ell_j^{\pm}$ can be obtained from the
three-body decay $\ell_i\to \ell_jl_k\bar{\ell}_k$, which receives
contributions from the $Z\ell_i \ell_j$ coupling via a virtual
$Z$. We will also consider the contribution from the $Z\ell_i
\ell_j$ and $W\ell_i\nu_{\ell_j}$ couplings to the one-loop decay
$\ell_i\to \ell_j\gamma$. In Sec. \ref{ndc}, a similar analysis is
performed within the nondecoupling scenario. We would like to
stress that, in contrast to what is observed in the decoupling
case, in the nondecoupling scenario the contributions from the
monopole and dipole structures may be equally important due to the
presence of strong-dynamics effects arising from the underlying
theory, thereby allowing two possible scenarios. In the first case
it is assumed that the monopole structure gives the dominant
contribution, which means that the most stringent bounds on $Z\to
\ell_i^{\mp}\ell_j^{\pm}$ can be obtained from the three-body
decay $\ell_i\to \ell_jl_k\bar{\ell}_k$. In the second scenario it
is assumed that the dipole contribution is the dominant one, which
implies that, due to the $SU_L(2)\times U_Y(1)$ symmetry, it is
possible to obtain bounds on $Z\to \ell_i^{\mp}\ell_j^{\pm}$ by using the
tree-level decays $\ell_i\to \ell_j\gamma$. It turns out that the
bounds obtained this way are the most stringent. Finally, we
present our conclusions in Sec. \ref{conc}.

\section{LFV in the decoupling scenario}
\label{dc}

In this section we assume that the underlying theory is of a decoupled nature. The
effective operators inducing LFV couplings were presented in a previous work
\cite{DT}. These operators can be classified  according to whether they induce the
$\gamma \ell_i \ell_j$ coupling or do not.

\subsection{Effective operators that only induce the $Z
\ell_i \ell_j$ vertex}

We can classify these operators in two classes. In the first place
we have those operators that can be generated at tree-level in a
fundamental theory. They are given by

\begin{mathletters}
\label{op1}
\begin{equation}
\label{op11}
{\cal O}_{\phi \ell}^{ij}=i\left( \phi^{\dagger
}D_\mu \phi \right) \left( \bar \ell_{Ri}\gamma^\mu
\ell_{Rj}\right),
\end{equation}
\begin{equation}
\label{op12}
{\cal O}_{\phi L}^{(1)ij}=i\left( \phi^{\dagger
}D_\mu \phi \right) \left( \bar L_i\gamma ^\mu L_j\right),
\end{equation}
\begin{equation}
\label{op13}
 {\cal O}_{\phi L}^{(3)ij}=i\left( \phi^{\dagger
}\tau^a D_\mu \phi \right) \left( \bar L_i\tau ^a\gamma^\mu
L_j\right) ,\nonumber
\end{equation}
\end{mathletters}

\noindent where $L_i$ and $\ell_{Ri}$ stand for the left-handed
doublet and the right-handed singlet of $SU_L(2)\times U_Y(1)$,
respectively, $\tau^a$ are the Pauli matrices and roman letter
indices are used to denote lepton flavors. The first two operators
induce the $Z\ell_i \ell_j$ and $H\ell_i \ell_j$ couplings,
whereas the third one also induces the $W\ell_i\nu_{\ell_j}$
vertex. Both the  $Z\ell_i \ell_j$ and $W\ell_i\nu_{\ell_j}$
couplings contribute to the one-loop induced decay $\ell_i\to
\ell_j\gamma$.

There is also another set of operators that can be generated at
the one-loop level or at a higher order:

\begin{mathletters}
\label{op2}
\begin{equation}
{\cal O}_{D\ell}^{ij}=\left(\bar L_i D_{\mu} \ell_{Rj}\right)
D^{\mu} \phi,
\end{equation}
\begin{equation}
{\cal O}_{DL}^{ij}=\left(\overline{D_\mu L_i} \ell_{Rj}\right)
D^{\mu} \phi.
\end{equation}
\end{mathletters}

Both of these sets of operators contribute to the three-body decay
$\ell_i \to \ell_j \ell_k \bar \ell_k$ via a virtual $Z$.

\subsection{Effective operators that induce both  the $Z \ell_i
\ell_j$ and $\gamma \ell_i \ell_j$ vertices}

Owing to gauge invariance, operators of this kind can only arise
at the one-loop level in any fundamental theory. According to the
Lorentz structure of these operators, we can classify them in two
categories:

\begin{mathletters}
\label{op3}
\begin{equation}
{\cal O}_{LW}^{ij}= ig \left(\bar L_i {\bf W}^{\mu \nu} \gamma_\mu
D_{\nu} L_j\right),
\end{equation}
\begin{equation}
{\cal O}_{LB}^{ij}= ig^\prime \left(\bar L_i B^{\mu \nu}
\gamma_\mu D_{\nu} L_j\right),
\end{equation}
\begin{equation}
{\cal O}_{\ell B}^{ij}= ig^\prime \left(\bar \ell_{Ri} B^{\mu \nu}
\gamma_\mu D_{\nu} \ell_{R j}\right),
\end{equation}
\end{mathletters}

\noindent and

\begin{mathletters}
\label{op4}
\begin{equation}
{\cal O}_{\ell W\phi}^{ij}= g \left(\bar L_i \sigma_{\mu \nu} {\bf
W}^{\mu \nu} \ell_{R j}\right) \phi,
\end{equation}
\begin{equation}
{\cal O}_{\ell B\phi}^{ij}= g^\prime \left(\bar L_i \sigma_{\mu \nu} B^{\mu \nu}
\ell_{R j}\right) \phi, \\
\end{equation}
\end{mathletters}

\noindent where ${\bf W}^{\mu \nu}= \tau^\alpha W^{\alpha \mu
\nu}$. It is understood that the hermitian conjugate of each
operator is to be added in the respective Lagrangian, i.e. ${\cal
L}_{\mathrm eff.} = (\alpha^{ij}/\Lambda^2) {\cal O}_{ij}+{\mathrm
H.c.}$ We have assumed that all the effective matrices $\alpha$
can not be simultaneously diagonalized by the unitary matrices
$V^\ell_L$ and $V^\ell_R$ that define the mass eigenstates. Note
that these groups of operators give rise to both $Z \ell_i \ell_j$
and $\gamma \ell_i \ell_j$ couplings as a direct consequence of
the $SU_L(2)\times U_Y(1)$ gauge invariance of the effective
theory. Therefore, the experimental constraints on $\ell_i \to
\ell_j \gamma$ can be easily translated into bounds on $Z \to
\ell^{\mp}_i \ell^{\pm}_j$. However, we will see below that these
operators play a marginal role in this decay, though the situation
may be different in the nondecoupling scenario.

\subsection{The most general $Z \ell_i \ell_j$ vertex and the decay $Z \to
\ell^{\mp}_i \ell^{\pm}_j$}

The effective operators shown in Eqs. (\ref{op1})-(\ref{op4})
induce the most general $Z \ell_i \ell_j$ vertex. In the case of
on-shell leptons, it is possible to make use of the Dirac equation
along with the Gordon identity to transform the Lorentz structure
induced by the operators of Eq. (\ref{op3}) into a dipole
structure. It turns out that the contribution from these operators
has terms that are proportional to $m_i/m_Z$ or $m_j/m_Z$, with
$m_{i,\,j}$ the lepton masses. It means that these operators give
a very suppressed contribution, as compared to that from the
operators of Eq. (\ref{op4}). Therefore, from now on we will not
consider the operators of Eq. (\ref{op3}). We thus can write the
most general structure of the $Z \ell_i \ell_j$ vertex in the
following way

\begin{equation}
\label{mzlg}
 {\cal M}^{Z\ell^i\ell^j}_\mu=\frac{ig}{2c_W} \bar{u}(p_i)\left[
\gamma_\mu\left(F_{1L}^{ij} P_L + F_{1R}^{ij} P_R\right) +
\frac{1}{m_Z} \left(F_{2L}^{ij} P_L + F_{2R}^{ij} P_R\right) k_\mu
+ \frac{i}{m_Z} F_{3R}^{ij} P_R \sigma_{\mu \nu} k^\nu
\right]v(p_j),
\end{equation}

\noindent where $k_\mu$ is the $Z$ four-momentum. We have defined
the following matrices in the flavor space

\begin{mathletters}
\begin{equation}
F_{1L}=-\left(\frac{v}{\Lambda}\right)^2
V^{\ell}_L\left(\alpha^{(1)}_{D\phi}+\alpha^{(3)}_{D\phi}\right)
V^{ \ell \dag}_L,
\end{equation}
\begin{equation}
F_{1R}=-\left(\frac{v}{\Lambda}\right)^2V^{\ell}_R\alpha_{D\phi}V^{\ell
\dag}_R,
\end{equation}
\end{mathletters}

\begin{mathletters}
\begin{equation}
F_{2L}= \frac{g^2}{2 \sqrt{2}
c_W^2}\left(\frac{v}{\Lambda}\right)^2V^{\ell}_L\alpha_{DL}V^{\ell
\dag}_R,
\end{equation}
\begin{equation}
F_{2R}= - \frac{g^2}{2 \sqrt{2}
c_W^2}\left(\frac{v}{\Lambda}\right)^2V^{\ell}_L\alpha_{D
\ell}V^{\ell \dag}_R,
\end{equation}
\end{mathletters}
\begin{equation}
F_{3R}=\frac{\sqrt{2}g}{c_W}\left(\frac{v}{\Lambda}\right)^2
V^{\ell}_L\left(c_W^2 \alpha_{{\ell}W\phi} + s_W^2 \alpha_{\ell
B\phi}\right)V^{{\ell} \dag}_R.
\end{equation}

\noindent It is evident that the terms proportional to $k_\mu$ in
Eq. (\ref{mzlg}) do not contribute when the $Z$ boson is on-shell.
We thus can conclude that the contributions to $Z \to \ell^{\mp}_i
\ell^{\pm}_j$ can only arise from the operators given in Eqs.
(\ref{op1}) and (\ref{op4}), i.e. only through the monopole and
dipole structures. Since the monopole structure can be generated
at tree-level by the underlying theory, its contribution will
dominate that from the dipole structure because the latter can
only arise at the one-loop level and has a suppression factor of
$(4\pi)^{-2}$. It is thus a good approximation to consider only
the contributions arising from the operators of Eq. (\ref{op1}).
In contrast, the $\gamma \ell_i\ell_j $ coupling is only induced
by the operators of Eqs. (\ref{op3}) and (\ref{op4}) since the
monopole contribution is forbidden because of electromagnetic
gauge invariance, i.e. the $\gamma \ell_i\ell_j $ coupling can
only arise at the one-loop level in any renormalizable theory.  In
order to obtain bounds on $Z \to \ell^{\mp}_i \ell^{\pm}_j$, we
will use the experimental bounds on the three-body decays
$\ell_i\to \ell_j \ell_k \bar{\ell_k}$, which may receive
contributions from the $Z\ell_i \ell_j$ coupling through a virtual
$Z$, mainly via the monopole structure. We will also calculate the
contributions from the  $Z\ell_i \ell_j$ and $W\ell_i\nu_{\ell_j}$
couplings to the one-loop induced decay $\ell_i\to \ell_j\gamma$
in order to analyze if this mode could be useful to obtain bounds
on $Z \to \ell_i^{\mp} \ell_j^{\pm}$. All these results can be
translated readily into the nondecoupling scenario, where the
dipole contribution to $Z \to \ell^{\mp}_i \ell^{\pm}_j$ will be
studied also. It turns out that, in that scenario, the dipole
contribution may be as important as that from the monopole
structure.

Taking into account just the contribution from the
tree-level-generated operators, the branching fraction for the
decay $Z \to \ell_i^{\mp} \ell_j^{\pm}$ can be written as

\begin{equation}
\label{brz} {\mathrm{BR}}\left(Z\to
\ell_i^{\mp}\ell_j^{\pm}\right) = \frac{\alpha}{3 s_{2W}^2}
\left(\frac{m_Z}{\Gamma_Z}\right) \left( |F_{1L}^{ij}|^2 +
|F_{1R}^{ij}|^2 \right),
\end{equation}

\noindent where we have neglected the lepton masses. We have also
introduced the definition $s_{2W}=2 c_W s_W$.

\subsection{Bounds from the three-body decay $\ell_i\to
\ell_j \ell_k\bar{\ell}_k$}

\begin{figure}[hbt]
\centerline{\epsfig{file=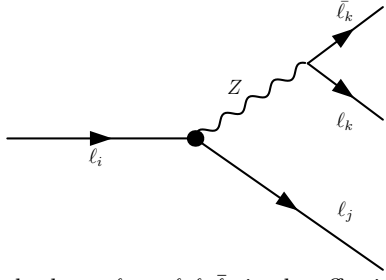,width=2in,clip=}}
\caption{Feynman diagram for the three-body decay $\ell_i\to
\ell_j \ell_k\bar{\ell}_k$ in the effective Lagrangian approach.
The dot denotes an effective LFV coupling.}
\label{liljlklk}
\end{figure}

The contribution from the $Z\ell_i \ell_j$ coupling to the decay
$\ell_i\to \ell_j \ell_k\bar{\ell}_k$ (viz Fig. \ref{liljlklk})
can be written as

\begin{equation}
\label{3br} {\mathrm{BR}}\left(\ell_i \to \ell_j \ell_k \bar
\ell_k\right) =\frac{a\, \alpha^2}{96 \pi s_{2W}^4}
\frac{m_i}{\Gamma_{\ell_i}} \left(\frac{m_i}{m_Z}\right)^4 \left(
|F_{1L}^{ij}|^2 + |F_{1R}^{ij}|^2 \right),
\end{equation}

\noindent with $a=1-4s^2_W+8s^4_W$ and $\Gamma_{\ell_i}$ being the full
$\ell_i$ width. Again we have neglected the final lepton masses,
i.e. $m_j=m_k=0$. From Eqs. (\ref{brz}) and (\ref{3br}) we can
obtain the following expression

\begin{equation}
\label{z3br} {{\mathrm{BR}}}\left(Z\to
\ell_i^{\mp}\ell_j^{\pm}\right) \le \frac{48 \pi s_{2W}^2}{a\,
\alpha} \left(\frac{\Gamma_{\ell_i}}{\Gamma_Z}\right)
\left(\frac{m_Z}{m_i}\right)^5 {\mathrm{BR}}_{{\mathrm
Exp.}}\left(\ell_i \to \ell_j \ell_k \bar{\ell}_k\right)
\end{equation}

\noindent where ${\mathrm{BR}}_{{\mathrm Exp.}}(\ell_i \to \ell_j
\ell_k \bar{\ell}_k)$ stands for the experimental constraints
\cite{pdg}:

\begin{mathletters}
\begin{equation}
{\mathrm{BR}}\left(\mu^- \to e^- e^- e^+\right)< 10^{-12},
\end{equation}
\begin{equation}
{\mathrm{BR}}\left(\tau^- \to \ell_j \ell_k
\bar{\ell_k}\right)<\kappa_{jk}\, 10^{-6},
\end{equation}
\end{mathletters}

\noindent and $\kappa_{jk}$ is a factor of order $O(1)$
corresponding to each different channel \cite{pdg}. These
equations allows us to obtain the following bounds

\begin{mathletters}
\label{b3}
\begin{equation}
{\mathrm{BR}}\left(Z \to \mu^{\mp} e^{\pm}\right)  \le  1.04
\times 10^{-12}
\end{equation}
\begin{equation}
{\mathrm{BR}}\left(Z \to \tau^{\mp} e^{\pm}\right)  \le  1.7
\times 10^{-5}
\end{equation}
\begin{equation}
{\mathrm{BR}}\left(Z \to \tau^{\mp} \mu^{\pm}\right)  \le   1.0
\times 10^{-5}.
\end{equation}
\end{mathletters}

\noindent These results are in agreement with those obtained from
unitarity-inspired arguments in Ref. \cite{nussinov}.

\subsection{Bounds from the two-body decay $\ell_i \to \ell_j\gamma$}

\begin{figure}[hbt]
\centerline{\epsfig{file=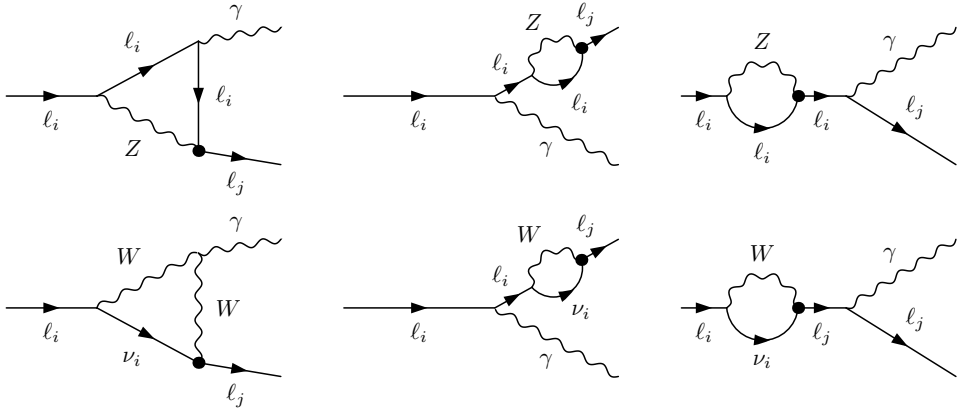,width=5in,clip=}}
\caption{Feynman diagrams for the radiative decay $\ell_i\to
\ell_j \gamma$ in the effective Lagrangian approach. The dot
denotes an effective LFV coupling. There is another set of
diagrams where the flavor-changing effective vertex is inserted in
the opposite end of the $Z$ boson or the neutrino.} \label{liljg}
\end{figure}

We now turn to study the contributions from the $Z\ell_i \ell_j$
and $W\ell_i\nu_{\ell_j}$ couplings to the one-loop decay
$\ell_i\to \ell_j\gamma$ (viz Fig. \ref{liljg}). While the
three-body decay $\ell_i\to \ell_j \ell_k \bar{\ell}_k$ gets
naturally suppressed by the three-body phase space and the
exchange of a virtual $Z$ boson, the one-loop decay $\ell_i\to
\ell_j\gamma$ gets a  suppression factor of $(4\pi)^{-2}$ plus an
extra power of $\alpha$. Since the current experimental
constraints on both decay modes are of the same order of
magnitude, the only way in which the radiative decay can compete
with the three-body decay is if the former arises from a
nondecoupling effect. However, we will see below that the
$\ell_i\to \ell_j\gamma$ amplitude is dominated by the virtual $Z$
and vanishes when $m_i/m_Z \to 0$.

\noindent The respective Feynman diagrams for the decay $\ell_i
\to \ell_j \gamma$ are shown in Fig. \ref{liljg}. We have used the
unitary gauge in our calculation. The expression for the $Z\ell_i
\ell_j$ coupling was given in Eq. (\ref{mzlg}), though we will
only consider the monopole contribution here. As for the
$W\ell_i\nu_{\ell_j}$ coupling, that can be induced by the
operators of Eq. (\ref{op13}), it is expressed
as

\begin{equation}
{\cal M}^{W\nu_i\ell_j}=\frac{ig\,\epsilon_L^{ij}}{\sqrt{2}}\,
\bar{u}(p_i) P_L\gamma_\mu v(p_j) W^\mu, \label{LWlilj}
\end{equation}

\noindent with

\begin{equation}
\epsilon_L=\left(\frac{v}{\Lambda}\right)^2 V^{\ell}_L
\alpha^{(3)}_{D\phi} V^{ \ell \dag}_L.
\end{equation}

After some calculation, the decay amplitude can be expressed as

\begin{equation}
{\cal M}^{\mu}\left(\ell_i \to \ell_j
\gamma\right)=\bar{u}(p_j)\left(f_V-f_A \gamma^5\right)\sigma^{\mu
\nu} q^\nu v(p_i),
\end{equation}

\noindent where $q$ is the photon four-momentum and the coefficients
$f_{V,A}$ are given as follows

\begin{equation}
f_{V,A}=\frac{\alpha}{4\pi s^2_{2W}}\left(A_R^{Vij}\pm
A_L^{Vij}\right),
\end{equation}

\noindent where the superscript $Vij$ denotes the contribution from the virtual boson ($Z$ or
$W$). As to the coefficients $A_{L\,R}^{Vij}$, they are given, in terms of scalar integrals,
by

\begin{eqnarray}
A_{L,R}^{Zij}&=&\frac{1}{4\, m_i^3\,
m_Z^2}\bigg[4\,m_Z^4\,g^{\ell}_{L,\,R}\left(
1 + B_0^1 + B_0^2 - B_0^3 - B_0^4 \right) \nonumber\\ &-&
2\,m_i^2\, m_Z^2\left(\pm 3\left(B_0^2-B_0^4\right) +
2\,g^{\ell}_{L,\,R} \left(B_0^1-B_0^3\right)\right) \nonumber\\
&-& m_i^4 \left(\pm 1-4\,g^{\ell}_{L,\,R}\,
m_Z^2\,C_0^1 \right)\bigg]F_{1L,\,1R}^{ij}, \label{A}
\end{eqnarray}

\noindent with $g_L^{\ell}=-1+2\,s_W^2$ and $g_R^{\ell}=2\,s_W^2$. By simplicity we
neglected the final lepton mass. The sign $+\,(-)$ holds for the $L\, (R)$ term.
As far as the $A_L^{Wij}$ coefficient is concerned, we have

\begin{equation}
A_L^{Wij}=\frac{2\,\epsilon_L^{ij}\,c_W^2}{m_i^3}\left[2\, m_W^2-
3\,m_i^2-2\,\left(m_W^2-m_i^2\right) \left(B_0^6 -
B_0^5+m_i^2\,C_0^2\right) \right],
\end{equation}

\noindent whereas $A_R^{Wij}=0$. The scalar integrals $B_0^i$ and $C_0^i$ are given, in the notation of Ref. \cite{LT}, as follows: $B_0^1=B_0(0,
m_Z^2, m_Z^2)$, $B_0^2=B_0(0, m_i^2, m_Z^2)$, $B_0^3=B_0(m_i^2, 0, m_Z^2)$,
$B_0^4=B_0(m_i^2, m_i^2, m_Z^2)$, $B_0^5=B_0(0, m_W^2, m_W^2)$, $B_0^6=B_0(m_i^2,
0, m_W^2)$, $C_0^1=C_0(m_i^2, 0, 0, m_i^2, m_Z^2, m_i^2)$ and $C_0^2=C_0(m_i^2, 0,
0, m_W^2, 0, m_W^2)$.

It is interesting to note that, although an effective vertex was
inserted into a one-loop diagram, from the above expressions it is
evident that the calculation renders a finite result. It can be
explained from the fact that the $Z\ell_i\ell_j$ coupling has a
renormalizable structure. Our result is very general in the sense that it can
also be applied to theories with an extra $Z^\prime$ boson with
LFV couplings of the monopole-structure form.

The branching ratio for the radiative decay $\ell_i \to \ell_j
\gamma$ is given by

\begin{equation}
{\mathrm{BR}}\left(\ell_i \to \ell_j
\gamma\right)=\frac{m^3_i}{8\pi\,\Gamma_{\ell_i}}
\left(|f_V|^2+|f_A|^2\right)=\frac{m^3_i\,\alpha^2}{64\,\pi^3\,\Gamma_{\ell_i}\,
s_{2W}^4} \left(|A_{L}^{Vij}|^2+ |A_{R}^{Vij}|^2\right)
\label{brt}.
\end{equation}

\noindent The scalar functions involved in the coefficients
$A_{L,R}^{Vij}$ can be numerically evaluated \cite{LT} or expanded
in powers of $m_i$. We will end up with an expression of the form
${\mathrm{BR}}\left(\ell_i \to \ell_j \gamma\right)=\beta_1
|F_{1L}^{ij}|^2+\beta_1 |F_{1R}^{ij}|^2+\beta_3
|\epsilon_L^{ij}|^2$, where the $\beta_k$ are some numerical
coefficients.

From the experimental side, we have

\begin{mathletters}
\label{cota2}
\begin{equation}
{\mathrm{BR}}\left(\mu \to e \gamma \right)<1.2 \times 10^{-11},
\end{equation}
\begin{equation}
{\mathrm{BR}}\left(\tau \to e \gamma \right)<2.9 \times 10^{-6},
\end{equation}
\begin{equation}
{\mathrm{BR}}\left(\tau \to \mu \gamma \right)<1.1 \times 10^{-6}.
\end{equation}
\end{mathletters}

\begin{figure}[hbt]
\centerline{\epsfig{file=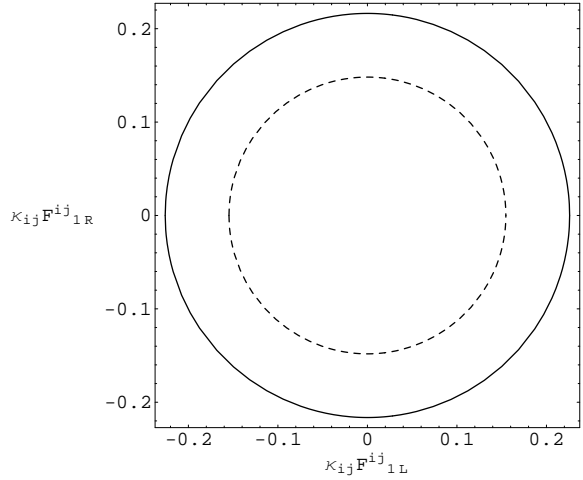,width=3.in,clip=}}
\caption{Bounds on the coefficients  $F_{1L,\,1R}^{ij}$: from $\mu
\to e \gamma$ (solid line) and $\tau \to \mu \gamma$ (dashed
line). The allowed region lies inside the curves. $\kappa_{\mu
e}=10^{-8}$ and $\kappa_{\tau \mu}=10^{-2}$.} \label{bounds}
\end{figure}

\noindent Therefore, from Eqs. (\ref{brt}) and (\ref{cota2}) we
can obtain an upper bound on the coefficients $F_{1L,\,1R}^{ij}$
and $\epsilon_L^{ij}$, which in turn can be used to put
constraints on the decay $Z \to \ell_i^{\mp} \ell_j^{\pm}$. In
Fig. \ref{bounds} we show the allowed region for the coefficients
$F_{1L,\,1R}^{ij}$, as obtained from the decays $\mu \to e \gamma$
and $\tau \to \mu \gamma$. As we are interested in obtaining upper
bounds on $F_{1L,\,1R}^{ij}$, we set $\epsilon_L^{ij}=0$. In the
plot of Fig. \ref{bounds}, the allowed regions, which
interestingly are almost circular in shape, lie inside the curves.
From these results and Eq. (\ref{brz}) we can obtain the following
bounds

\begin{mathletters}
\label{boundsliljg}
\begin{equation}
{\mathrm{BR}}\left(Z\to \mu^{\pm} e^{\pm} \right)\le6.12 \times
10^{-11},
\end{equation}
\begin{equation}
{\mathrm{BR}}\left(Z \to \tau^{\pm} \ell^{\pm} \right)\le2.8\times10^{-5},
\end{equation}
\end{mathletters}

\noindent where $\ell=e $ or $\mu$. Although these bounds are
weaker than the ones obtained from the three-body decay $\ell_i\to
\ell_j \ell_k \bar{\ell_k}$, they show that the one-loop decay
$\ell_i \to \ell_j \gamma$ may also be useful to obtain bounds on
$Z \to \ell_i^{\mp} \ell_j^{\pm}$.

\section{LVF in the nondecoupling scenario}
\label{ndc}

In the scenario where the underlying new physics effects arise
from a strongly interacting sector, the relevant LFV operators are
similar to those given in the decoupling scenario, but now with
the Higgs doublet replaced by the following unitary matrix

\begin{equation}
U=\exp\left(\frac{2i\tau^a\varphi^a}{v}\right),
\end{equation}

\noindent where $\varphi^a$ stands for the Goldstone bosons. In
this realization of the $SU_L(2)\times U_Y(1)$ group, the
covariant derivative is defined as ${\bf D}_\mu U=\partial_\mu
U+ig {\bf W}_\mu U-ig'U{\bf B}_\mu$, with the Abelian field given
by ${\bf B}_\mu=(\tau^3/2)B_\mu$. From the discussion presented
before, it is clear that the relevant operators  are the analogous
of those shown in Eqs. (\ref{op1}) and (\ref{op4}), although the
operator ${\cal O}^{(3)ij}_{\phi L}$ has no nonlinear counterpart.
These operators can be written as

\begin{mathletters}
\begin{equation}
{\cal L}_{UR}=i\lambda^{ij}_{UR}{\mathrm Tr}\left[\tau^3U^\dag
{\bf D}_\mu U\right]\bar{R}_i\gamma^\mu R_j+{\mathrm
H.c.},\label{op51}
\end{equation}
\begin{equation}
{\cal L}_{UL}=i\lambda^{ij}_{UL}{\mathrm Tr}\left[\tau^3U^\dag
{\bf D}_\mu U\right]\bar{L}_i\gamma^\mu L_j+{\mathrm
H.c.},\label{op52}
\end{equation}
\end{mathletters}

\begin{mathletters}
\begin{equation}
{\cal L}_{\ell WU}=g\frac{\lambda^{ij}_{\ell
WU}}{\Lambda}\left(\bar{L}_i\sigma_{\mu \nu }{\bf W}^{\mu \nu
}UR_j\right)+{\mathrm H.c.},\label{op61}
\end{equation}
\begin{equation}
{\cal
L}_{lBU}=g'\frac{\lambda^{ij}_{lBU}}{\Lambda}\left(\bar{L}_i\sigma_{\mu
\nu }UR_j\right)B^{\mu \nu }+{\mathrm H.c.}, \label{op62}
\end{equation}
\end{mathletters}

\noindent where $R_i=(0,\ell_{Ri})$. In this scenario there is an
upper bound on the new physics scale, i.e. $\Lambda \sim 4\pi v$,
which will adopted below. Notice that the first group of operators
have dimension 4 in mass units, whereas the last ones have
dimension 5.

The formulas given in Eqs.(\ref{mzlg})-(\ref{brz}) also hold, but the $F_i$
matrices are to be replaced by

\begin{mathletters}
\begin{equation}
A_{1L}=-2V^\ell_L\,\lambda_{UL}\,V^{\ell\dag}_L,
\end{equation}
\begin{equation}
A_{1R}=-2V^\ell_R\,\lambda_{UR}\,V^{\ell\dag}_R,
\end{equation}
\begin{equation}
A=\frac{g^2}{2\pi c^2_W}V^\ell_L\left(c^2_W\lambda_{\ell
WU}-s^2_W\lambda_{\ell BU}\right)V^{\ell \dag}_R.
\end{equation}
\end{mathletters}

Since we are assuming a strong-interaction as responsible for the
LFV effects, two scenarios are of interest. In the first case we
will assume that the monopole contribution dominates over the
dipole contribution, whereas in the second case we will take the
dipole moment contribution as being the dominant one.

\subsection{Monopole dominance}

In this scenario it is assumed that the structure induced by the
operators (\ref{op51}) and (\ref{op52}) gives the dominant
contribution. From the above discussion, it is clear that the most
stringent bounds can be obtained from the three-body decay
$\ell_i\to \ell_j \ell_k\bar{\ell}_k$. It is also clear that the
relation given in Eq. (\ref{z3br}) still holds. Consequently, the
respective bounds are the same as those given in Eqs. (\ref{b3}).
Finally, the bounds arising from the decay $\ell_i \to \ell_j
\gamma$ also hold.

\subsection{Dipole moment dominance}

We now neglect the monopole term and focus on the contribution
arising from the operators (\ref{op61}) and (\ref{op62}). Due to
$SU_L(2)\times U_Y(1)$ gauge invariance, these operators induce
both the $Z\ell_i \ell_j$ and $\gamma \ell_i \ell_j$ vertices,
which means that the decay $\ell_i\to \ell_j\gamma$ can give more
stringent bounds than the ones arising from the three-body decay
$\ell_i\to \ell_j \ell_k\bar{\ell}_k$. This is a consequence that
the electromagnetic decay has a phase space factor less restricted
than the three-body decay and does not involve the factor
$(m_i/m_Z)^4$ coming from the inclusion of the virtual $Z$ but
only the kinematic one $(m_i/m_Z)^2$.

The branching fractions for the decays $Z\to
\ell_i^{\mp}\ell_j^{\pm}$ and $\ell_i\to \ell_j\gamma$ can now be
written as

\begin{equation}
{\mathrm{BR}}\left(Z\to \ell_i^{\pm}\ell_j^{\pm}\right)=
\frac{\alpha}{6s^2_{2W}}\left(\frac{m_Z}{\Gamma_Z}\right)|A^{ij}|^2,
\end{equation}

\begin{equation}
{\mathrm{BR}}(\ell_i\to\ell_j\gamma)=\frac{3\alpha}{4s^2_{2w}}
\left(\frac{m_i}{\Gamma_{\ell_i}}\right)\left(\frac{m_i
}{m_Z}\right)^2|B^{ij}|^2,
\end{equation}

\noindent where

\begin{equation}
B=\frac{e}{2\pi
c^2_W}V^\ell_L\left(\lambda_{\ell BU}+\lambda_{\ell WU}\right)V^{\ell\dag}_R.
\end{equation}

After introducing the experimental constraints on the
electromagnetic decays, we have

\begin{equation}
{\mathrm{BR}}(Z \to \ell_i^{\mp} \ell_j^{\pm}) \le \frac{2}{9}
\left(\frac{\Gamma_{\ell_i}}{\Gamma_Z}\right)\left(\frac{m_Z}{m_i}\right)^3
{\mathrm{BR}}_{\mathrm Exp.}\left(\ell_i \to \ell_j \gamma\right)
\left\{
\begin{array}{c}
\left(\frac{c_W}{s_W}\right)^2 \qquad {\mathrm for} \quad {\mathcal L}_{\ell WU}^{ij} \\
\left(\frac{s_W}{c_W}\right)^2 \qquad {\mathrm for}\quad {\mathcal
L}_{\ell BU}^{ij},
\end{array}\right.
\end{equation}

\noindent By using the respective experimental constraints
\cite{pdg} we get

\begin{mathletters}
\label{boundnd}
\begin{equation}
{\mathrm{BR}}(Z \to \mu^{\mp} e^{\pm}) \le  \left(8.64 \times
10^{-22},\,7.81 \times 10^{-23}\right),
\end{equation}
\begin{equation}
{\mathrm{BR}}(Z \to \tau^{\mp} e^{\pm}) \le \left(3.09 \times
10^{-13},\,2.79 \times 10^{-14}\right),
\end{equation}
\begin{equation}
{\mathrm{BR}}(Z \to \tau^{\mp} \mu^{\pm}) \le \left(1.25 \times
10^{-12},\,1.13 \times 10^{-13}\right).
\end{equation}
\end{mathletters}

\noindent where the first (second) figure in the parenthesis
correspond to the operator ${\cal L}_{\ell WU}$ (${\cal L}_{\ell
BU}$). It should be noticed that the same bounds apply in the
decoupling case, in the unlikely scenario where the dipole
contribution dominates over that from the monopole. The above
bounds have severe consequences. They imply that the existing
experimental constraints on the decays $\ell_i\to \ell_j\gamma$,
together with $SU_L(2)\times U_Y(1)$ gauge invariance, are enough
to rule out any possible detection of a LFV transition of the $Z$
boson if it arises via a $Z\ell_i \ell_j$ coupling of the form of
a dipole moment.

\section{Final discussion}
\label{conc}

Until now, the LFV decay $Z \to \ell_i^{\mp} \ell_j^{\pm}$ have
been studied within a large variety of models
\cite{illana,iltan,ghosal,langacker,ZLFV}. These studies show
that, at least for some values of the model parameters, the
respective decay rates might be at the reach of the planned TESLA
collider. However, all of these analyses rely on several
assumptions about the parameters of the model under study. We have
shown in this work that an ELA analysis is well suited for
studying this LFV decay. We have considered both the linear and
nonlinear realizations of the ELA. This approach has allowed us to
make some general predictions about the $Z \to \ell_i^{\mp}
\ell_j^{\pm}$ rates starting from the current experimental bounds
on the low energy processes $\ell_i\to \ell_jl_k\bar{\ell}_k$ and
$\ell_i\to \ell_j\gamma$. We summarize our results in Table
I. We also analyzed the impact of the LFV couplings
$Z\ell_i \ell_j$ and $W\ell_i\nu_{\ell_j}$ on the muon anomalous
magnetic moment $(g-2)_\mu$, but our calculation showed that the
bounds obtained this way are rather weak. So, we refrain from
showing the respective results here. In the same context, there
are other process that could be useful to obtain constraints on
the LFV $Z$ boson couplings, such as $\mu-e$ conversion and
muonium-antimuonium conversion. We preferred the decays $\ell_i\to
\ell_jl_k\bar{\ell}_k$ and $\ell_i\to \ell_j\gamma$ since they do
not imply any extra assumption.

In this work we have examined some potential scenarios for the
contributions arising from the two Lorentz structures associated
with the on-shell $Z\ell_i \ell_j$ vertex, namely the monopole and
dipole terms. It was shown that, in the decoupling scenario, the
decay $Z \to \ell_i^{\mp} \ell_j^{\pm}$ arises mainly from the
monopole term. In this case the strongest constraints on these
processes are obtained from the current bounds on the decays
$\ell_i\to \ell_jl_k\bar{\ell}_k$, though the constraints on
$\ell_i\to \ell_j\gamma$ are also useful for the same purpose. On
the other hand, in the nondecoupling scenario, where the LFV
effects have a strongly interacting origin, gauge invariance as
the main ingredient of the effective theory induce simultaneously
both $Z\ell_i \ell_j$ and $\gamma \ell_i \ell_j$ vertices. In this
scenario it might be that both the monopole and dipole contributions
have the same strength. It happens that if the main contribution
came from the dipole term, the current limits on the decay $\ell_i
\to \ell_j \gamma$ would place severe constraints on $Z \to
\ell_i^{\mp} \ell_j^{\pm}$ [see Eqs. (\ref{boundnd})], which
clearly are far from the reach of the planned TESLA collider.
These results suggest indeed that the dipole contribution is
unlikely to be observed.

In summary, if the new physics LFV effects are of decoupled nature,
the most stringent bound on $Z\to \mu^{\mp} e^{\pm}$ is of the
order of $10^{-12}$, which suggests that this mode would be out of
the reach of TESLA. Since the current experimental limits on
$\tau\to \ell_jl_k\bar{\ell}_k$ are less stringent than that on
$\mu^- \to e^- e^- e^+$, the resulting bounds on$Z\to \tau^{\mp}
\ell^{\pm}$ are also weaker than that on $Z\to \mu^{\mp} e^{\pm}$.
As a consequence, the decay $Z\to \tau^{\mp} \ell^{\pm}$ may still
be at the reach of the TESLA collider. In this respect, it has
been conjectured that LFV effects might be more evident in
transitions involving the $\tau$ lepton.

\acknowledgments{ Support from CONACYT and SNI (M\' exico) is
acknowledged. JJT and JMH thank J. L. D\'\i az-Cruz for his
comments.}

\newpage

\begin{table}
\label{table}
\begin{tabular}{cccc}
&$\ell_i\to\ell_j \ell_k\bar{\ell}_k$&$\ell_i\to
\ell_j \gamma$ (one-loop level)&$\ell_i\to\ell_j \gamma$ (tree-level)\\
\hline
${\mathrm{BR}}\left(Z \to \mu^{\mp} e^{\pm}\right)$&$\le  1.04
\times 10^{-12}$&$\le 6.12 \times 10^{-11}$&
$\le \left(10^{-22}-\times 10^{-23}\right)$\\
\hline
${\mathrm{BR}}\left(Z \to \tau^{\mp} e^{\pm}\right)$&$\le  1.7
\times 10^{-5}$&$\le\sim 10^{-5}$&$\le \left(10^{-13}-\times 10^{-14}\right)$\\
\hline
${\mathrm{BR}}\left(Z \to \tau^{\mp} \mu^{\pm}\right)$&$\le   1.0
\times 10^{-5}$&$\le\sim 10^{-5}$&$\le \left(10^{-12}-\times 10^{-13}\right)$
\end{tabular}
\caption{Constraints on the LFV decays $Z\to \ell^{\mp}_i\ell^{\pm}_j$ as obtained
from the experimental bounds on  $\ell_i\to
\ell_j \ell_k\bar{\ell}_k$ and $\ell_i\to
\ell_j \gamma$. The third column correspond to the monopole term of the
$Z\ell_i\ell_j$ coupling, which can induce the decay $\ell_i\to
\ell_j \gamma$ at the one-loop level. The last column
is obtained in the scenario where the new physics LFV effects only contribute to the
dipole term of $Z\ell_i\ell_j$. The operators that induce this term can also give rise
to $\ell_i\to\ell_j \gamma$ at tree-level.}
\end{table}

\end{document}